\documentclass[%
5p,hyperref,sort&compress,
 reprint,
 amsmath,amssymb,
 aps,
]{revtex4-2}
\usepackage[
pdfauthor={derajan},
pdftitle={How to do this},
pdfstartview=XYZ,
bookmarks=true,
colorlinks=true,
linkcolor=blue,
urlcolor=blue,
citecolor=blue,
pdftex,
bookmarks=true,
linktocpage=true,   
hyperindex=true
]{hyperref}

\usepackage {braket}
\usepackage{graphicx}
\usepackage{dcolumn}
\usepackage{float}
\usepackage{bm}
\usepackage{overpic}
\usepackage{amsmath}
 \usepackage{multirow}

\begin{document}

\preprint{APS/123-QED}

\title{Practical No-Switching Continuous-Variable Quantum Key Distribution with Biased Quadrature Detection}

\author{Jiale Mi} \author{Yiming Bian} \author{Lu Fan} \author{Song Yu} \author{Yichen Zhang} \email{zhangyc@bupt.edu.cn} 

\affiliation{%
State Key Laboratory of Information Photonics and Optical Communications, School of Electronic Engineering, Beijing University of Posts and Telecommunications, Beijing {100876}, China
}%
\date{\today}
\begin{abstract}
Continuous-variable quantum key distribution protocol using coherent states and heterodyne detection, called No-Switching protocol, is widely used in practical systems due to the simple experimental setup without basis switching and easy assessment to phase information. 
The security of an ideal No-Switching protocol has been proved against general attacks in finite-size regime and composable security framework, whose heterodyne detector consists of a beam splitter with transmittance of $50\%$ and two ideal homodyne detectors. However, the transmittance of a beam splitter is inaccurate and the two detectors always have different quantum efficiency and electronic noise, which introduce asymmetry into the heterodyne detection, and further lead to the mismatch between the ideal protocol and practical systems, thereby overestimating the secret key rate and resulting in a practical security loophole. 
In this paper, we close this loophole by proposing a modified No-Switching protocol with biased quadrature detection, where the asymmetry of the heterodyne detection is modeled to match the practical systems, and the security of the protocol is analyzed in asymptotic and finite-size regimes.
Further, an optimization strategy is proposed to achieve the optimal secret key rate by adjusting the transmittance of the beam splitter. Simulation results show the necessity of considering the asymmetry in heterodyne detection and the effectiveness of the optimization, which provides a promising way to realize a practical secure and high-performance No-Switching system.
\end{abstract}

\maketitle

\section{\label{sec:level1}Introduction}
Quantum key distribution (QKD) \cite{bennett1984update} stands out among many secure communication technologies by guaranteeing its unconditional security based on quantum uncertainty principle and quantum non-cloning theorem \cite{pirandola2020advances,xu2020secure}. In the mainstream QKD protocol framework, there are two categories based on the different encoding dimensions at the source: discrete variable quantum key distribution (DV-QKD) and continuous variable quantum key distribution (CV-QKD) \cite{weedbrook2012gaussian,diamanti2015distributing}. 
CV-QKD distributes secret keys in an infinite-dimensional Hilbert space with the quadrature of optical field as information carrier \cite{grosshans2002continuous,weedbrook2004quantum}. The protocols using coherent states have been widely applied in practical experiments \cite{zhang2023continuous}, which are simple to implement and have higher compatibility with the existing communication components \cite{guo2021toward}. Recent experimental achievements present the potential for the long-distance transmission within metropolitan areas \cite{jouguet2013experimental,zhang2019continuous,wang2020high,zhang2020long,wang2022sub} and large-scale application \cite{zhang2019integrated,eriksson2019wavelength,bian2023high,hajomer2023continuous}, making them a promising way of future secure communications systems.
 
CV-QKD using Gaussian-modulation coherent state with heterodyne detection (No-Switching) protocol \cite{weedbrook2004quantum} has been widely used in practical applications, which enables the key distribution without the need for random basis switching, leading to simple experimental setup and easy assessment to phase information \cite{weedbrook2006coherent}. The security of No-Switching protocol has been extensively demonstrated and can resist arbitrary attacks in the asymptotic regime, as well as the case of  finite-size regime and composable security framework \cite{grosshans2005collectiveattacks,pirandola2009direct,renner2009finetti,leverrier2013security,leverrier2015composable,leverrier2017security}. Relevant practical experiments and field tests have also proved its feasibility \cite{lance2005no,symul2007experimental,brunner2017low,kleis2017continuous,jain2022practical,pi2023sub}. No-Switching protocol is attracting further exploration by researchers for its advantages in terms of cost, performance and integration \cite{pietri2023cv,aldama2023inp}.

No-Switching protocol using heterodyne detection is widely adopted by the mainstream systems nowadays \cite{jain2022practical,pi2023sub} in order to avoid the selection of the measurement basis at the receiving end and simplify the implementation of the experiment. An ideal heterodyne detection model consists of a beam splitter with transmittance of 50\% and two ideal homodyne detectors. However, in the practical system, the imperfections of detector and other devices often introduce asymmetry into the heterodyne detection, which makes the original No-Switching protocol limited in practical application. More specifically, the beam splitter cannot be completely symmetrical with the transmittance of $50:50$, detectors also have inevitable imperfecftions: limited detection efficiency and electronic noise \cite{lodewyck2007quantum}, which is proved by the relevant measurement of experimental devices. The different imperfections often lead to an asymmetric effect to the practical systems, thus reducing the secret key rate. 
Moreover, the neglect of the asymmetry in practical experiments will lead to an inaccurate parameter estimation, which introduces a practical security loophole to the system. 

In this article, we present the mismatch between the existing entanglement-based (EB) model of ideal protocol and the prepare-and-measure (PM) model of practical system caused by the asymmetry in heterodyne detection, which threatens the practical security of the CV-QKD protocols. Therefore, we propose a biased No-Switching protocol, which can well reflect the imperfections of the current practical systems and bridge the gap between the EB model and the PM model, thus achieving more secure implementation of the protocol and making the protocol more in line with the needs of practical applications.
In this biased No-Switching protocol, we apply a modified heterodyne detector model \cite{zhang2020one} and verify its security and feasibility. Furthermore, by selecting the optimal transmittance of the beam splitter at different distances, we can reduce the deviation caused by detector asymmetry, achieve a higher secret key rate and improve the performance of the protocol. 
Simulation results comparing the performance of the ideal and biased protocols are provided to show the necessity of the biased No-Switching protocol for practical systems.

The rest of the paper is organized as follows. In Sec.~\ref{sec:level2}, we present the problem of mismatch between the ideal EB model and the practical PM model, as well as some experimental data to prove the asymmetry in heterodyne detection. Next in Sec.~\ref{sec:level3}, a new biased No-Switching protocol is proposed with its security analysis given in detail. Sec.~\ref{sec:level4} shows multiple simulation results and discussion in varies cases. Finally conclusions are given in Sec.~\ref{sec:level5}.

\section{\label{sec:level2}Mismatch Between the Ideal No-Switching Protocol and the Practical System}
In this section, we mainly present the problem that the EB model of the ideal protocol does not match the PM model of the practical experiment, with the process of trusted detector modeling and security analyses theoretically, as well as the experimental proof practically. 

\subsection{The Ideal Entanglement-Based Model and the Practical Prepare-and-Measurement Model}
\begin{figure*}[t]
\includegraphics[width=16cm]{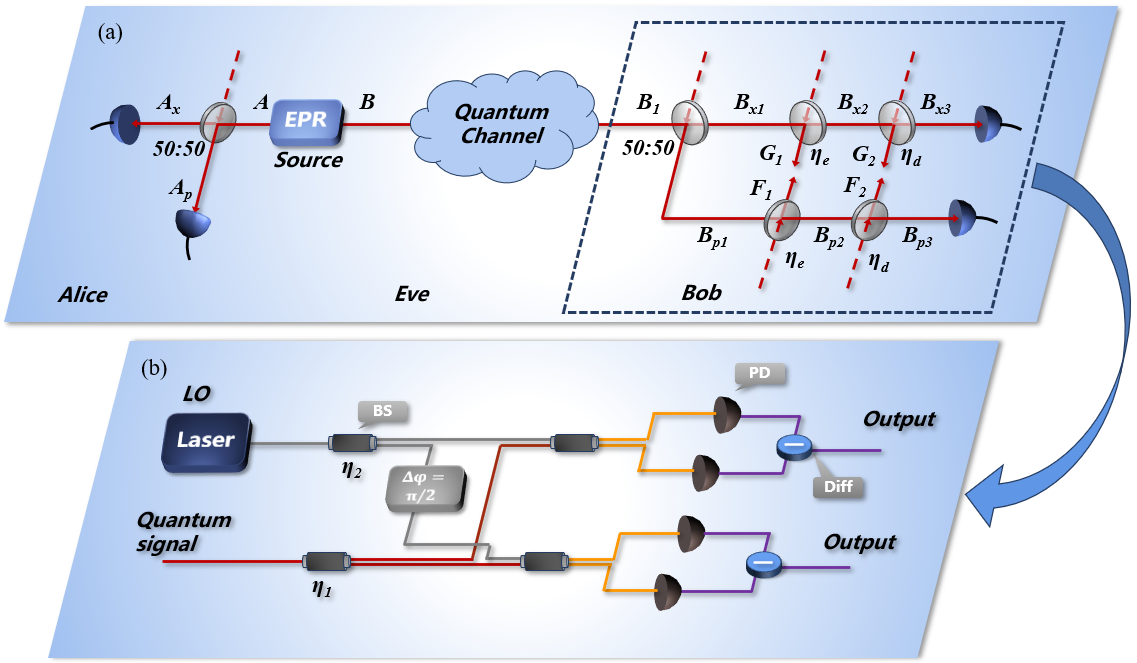}
\caption{\label{fig1} (\textbf{a}) The entanglement-based scheme of the ideal No-Switching protocol with a modified detector model, where Alice keeps one of the EPR states and carries out heterodyne detection while the other mode is sent through the quantum channel and performed heterodyne detection by Bob. The heterodyne detector is modeled by a beam splitter with transmittance $50:50$ and two identical homodyne detectors, where $\eta_d$ represents the detecting efficiency of the two detectors and $\eta_e$ represents the electronic noise. (\textbf{b}) The prepare-and-measure scheme of the practical biased heterodyne detection for No-Switching protocol. For the heterodyne detection, the quantum and local oscillator are divided by two beam splitters with different transmittance $\eta_1$ and $\eta_2$ respectively. A phase shift of $\pi/2$ between the two local oscillators is introduced to realize the detection of $x$ and $p$ quadratures of the quantum signals. LO: local oscillator, BS: beam splitter, PD: photodiode, Diff: differentiator.}
\end{figure*}

The EB scheme of the ideal No-Switching protocol is shown in Fig.~\ref{fig1} (\textbf{a}), in which a modified detector modeling method achieving one-time calibration is chosen. This model has been widely used in the current study, where the electronic noise of the detector is described by a beam splitter with transmittance $\eta_{e}$, while the detection efficiency is described by another beam splitter with transmittance $\eta_{d}$. First of all,  Alice prepares $n$ EPR states with the variance of $V_A$. She keeps one mode $A$, and sends the other mode $B$ through the quantum channel to Bob. The mode $B_1$ from the quantum channel is split into two modes, $B_{x_1}$ and $B_{p_1}$, after passing through a beam splitter with the splitting ratio of $50:50$. Alice and Bob then perform heterodyne detection on the reserved mode and the received mode respectively. Finally, Alice and Bob utilize the corrected data to perform parameter estimation, post-processing, data reconciliation, and privacy amplification to obtain the final secret key. 

The EB scheme mentioned above is primarily used for security analysis and the calculation of secret key rate, while in practical experiments, the PM model is always chosen, which is shown in Fig.~\ref{fig1} (\textbf{b}). Here we only present the detection part at Bob's side in detail for the subsequent analysis. It is obvious from the PM model that the heterodyne detector is realized by combining two homodyne detectors together. Before the detection, the quantum and local oscillator (LO) signals are first divided by two beam splitters with the transmittance of $\eta_1$ and $\eta_2$ respectively, and a phase shift of $\pi/2$ between the two LOs is introduced to realize the detection of $x$ and $p$ quadratures of the quantum signals. Then the two sets of signals are detected by the homodyne detectors separately, where for each of them, the quantum and LO signals are interfered by a coupler, and then two photodiodes are used to detect the two output signals. The two branches of the photocurrent generated by the photodiode are differentiated and output as the final step of the detection process. 

The practical security of the CV-QKD protocols relies on the equivalence between the PM model and the EB model, where the output of the PM model should be normalized by shot noise unit (SNU). We first write the output of the EB model as follows:
\begin{equation}
		\hat{x}_{hom}=\sqrt{\eta_e}(\sqrt{\eta_d}\hat{x}_{B}+\sqrt{1-\eta_d}\hat{x}_{v1})+\sqrt{1-\eta_e}\hat{x}_{v2},
\end{equation}
where $\hat{x}_{v1}$, $\hat{x}_{v2}$ represents the vacuum coupled in by the two beam splitters, both of which are Gaussian variables with a variance of 1.

Corresponding to this model is a new shot noise unit under the new trusted noise modeling. In the PM model, the method to obtain SNU is: keep the local oscillator path on, the signal optical path off, and calculate the variance of the output as the shot noise, which is defined as $u_s^{OTC}=A^2X_{LO}^2+V_{el}$ \cite{huang2020modified,chu2021practical,huang2021continuous}, and $V_{el}$ is the variance of $X_{ele}$. Using this new shot noise unit, the actual output of the detector can be normalized as follows:
\begin{equation}
\begin{aligned} 
x_{out}^{new}=\frac{AX_{LO}}{\sqrt{A^2X_{LO}^2+V_{el}}}(\sqrt{\eta_d}\hat{x}_{B}+\sqrt{1-\eta_d}\hat{x}_{v1})+ \\
\frac{\sqrt{V_{el}}}{\sqrt{A^2X_{LO}^2+V_{el}}}\hat{x}_{v2}.\\
\end{aligned}
\end{equation}

At this point, if we assume $\eta_e=A^2X_{LO}^2/(A^2X_{LO}^2+V_{el})$, we can get the result that the practical detector output is equivalent to the modified practical detector output, and $x_{out}^{new}=\hat{x}_{hom}$. Therefore, the security and feasibility are proved. Furthermore, if we want to get the relationship between the different models of electronic noise in the conventional model and the modified model, we can normalize $V_{el}$ with conventional SNU and replace it with $v_{el}$ to get 
\begin{equation}
		\eta_e=\frac{1}{1+v_{el}}.
	\end{equation}
	
According to the EB module of the ideal No-Switching protocol, the secret key rate can be expressed as:
\begin{equation}
		R=\beta I_{AB}-S_{BE},\label{(4)}
\end{equation}
where $\beta$ represents the reconciliation efficiency \cite{van2004reconciliation,milicevic2018quasi,ma2023practical}, $I_{AB}$ is the mutual information between Alice and Bob, and $S_{BE}$ is the mutual information between Eve and Bob. The mutual information between Alice and Bob can be seen as the sum of the mutual information obtained from two homodyne detectors, which is described by the Shannon entropy, 
	\begin{equation}
		I_{AB}=2 \times \frac{1}{2}\log_2\frac{V_B}{V_{B|A}}=\log_2\frac{V+\chi_{tot}}{1+\chi_{tot}},
	\end{equation}
where $\chi_{tot}$ is defined as $\chi_{tot}=\chi_{line}+\chi_{het}/T$ for conciseness. The excess noise introduced by Gaussian channels is defined as $\chi_{line}=1/T-1+\varepsilon$, and the detection-added noise referred to Bob's side is defined as $\chi_{het}=[1+(1-\eta_d)+2v_{el}]/\eta_d$, the channel transmissivity $T=10^{-\alpha L/10}$, where the fiber loss coefficient $\alpha$ = 0.2 dB/km, $L$ represents the transmission distance.

The mutual information between Eve and Bob is given by the Holevo bound of their von Neumann entropy $\chi_{BE}$,
	
\begin{equation}
	\begin{aligned}
		\chi_{BE}&=S(\rho_{AG_2B_3})-S(\rho_{AG_2}^{p_B,x_B}) \\
  &=\sum_{i=1}^3G(\frac{\lambda_i-1}{2})-\sum_{i=4}^{5}G(\frac{\lambda_i-1}{2}),
		\end{aligned}
\end{equation}
here $\lambda_{1\sim3}$ is the symplectic eigenvalue of the covariance matrix $\gamma_{AG_2B_3}$, which is written as
\begin{equation}
    \gamma_{AG_2B_3}=
\begin{bmatrix}
\gamma_{AG_2} & \sigma_{AG_2B_3}^T \\
\sigma_{AG_2B_3} & \gamma_{B_3}
\end{bmatrix}. \label{7}
\end{equation}
$\lambda_{4\sim5}$ is the symplectic eigenvalue of the covariance matrix $\gamma_{AG_2}^{p_B,x_B}$, which can be derived from the covariance matrix $\gamma_{AG_2B_3}$:
\begin{equation}
\gamma_{AG_2}^{p_B,x_B}=\gamma_{AG_2}-\sigma_{AG_2B_3}^TH_{het}\sigma_{AG_2B_3},
\end{equation}
and $H_{het}=(\gamma_{B_3}+I_2)^{-1}$, $I_2=
\begin{bmatrix}
		1 & 0 \\
		0 & 1
\end{bmatrix}$. The other matrices have also been obtained from Eq.~\eqref{7}, so that we can calculate the secret key rate.

For the theoretical security analysis mentioned above, the detection efficiency of the detector as well as other channel parameters are calibrated by the results in practical experiments, where the beam splitters, couplers and photodiodes always have imperfections, which results in the asymmetry in the heterodyne detection. To be specific, the transmittance of the beam splitters for the quantum and LO signals $\eta_1$ and $\eta_2$ marked in the PM model are not precisely $50:50$, and the detection efficiency and electronic noise of the two homodyne detectors are not exactly the same, which will introduce asymmetry to the heterodyne detection. However, in theoretical security analysis, the heterodyne detector is generally modeled as a $50:50$ beam splitter and two identical homodyne detectors, where their detection efficiency and electronic noise are exactly the same. Therefore, the existing EB model in Fig.~\ref{fig1} does not match the PM model reflecting the practical experiment, which threatens the practical security of the No-Switching protocol.

\subsection{Experimental Results of Key Devices of Heterodyne Detection}
In this part, the source of asymmetry in heterodyne detection is presented through practical measurements. We mainly measure the actual beam splitting ratio of a $50:50$ beam splitter and compare the output noise power of two pairs of homodyne detectors with the same input LO power respectively.

In the experiment, a new beam splitter with low insertion loss is chosen to make the measurement results more reliable. The beam splitter has two input ports and two output ports, we connect the quantum signal light and local oscillator light respectively to the two input ports, and then measure the optical power of the two output ports. Five representative sets of data are provided, and the splitting ratio for each set is calculated respectively. The results are listed in Table~\ref{tab1} and Table~\ref{tab2}. Through these data, we can see that the beam splitting ratio of the beam splitter is not precisely $50:50$, there are always errors in each measurement result, some of which can be quite significant and will do great harm to the experiment. The insertion loss of both output ports is not exactly the same with different input powers. Therefore, a practical beam splitter always has insertion loss and inaccurate splitter ratio, which will bring problems to our experiments and introduce security loopholes.

\begin{figure*}[t]
\includegraphics[width=16cm]{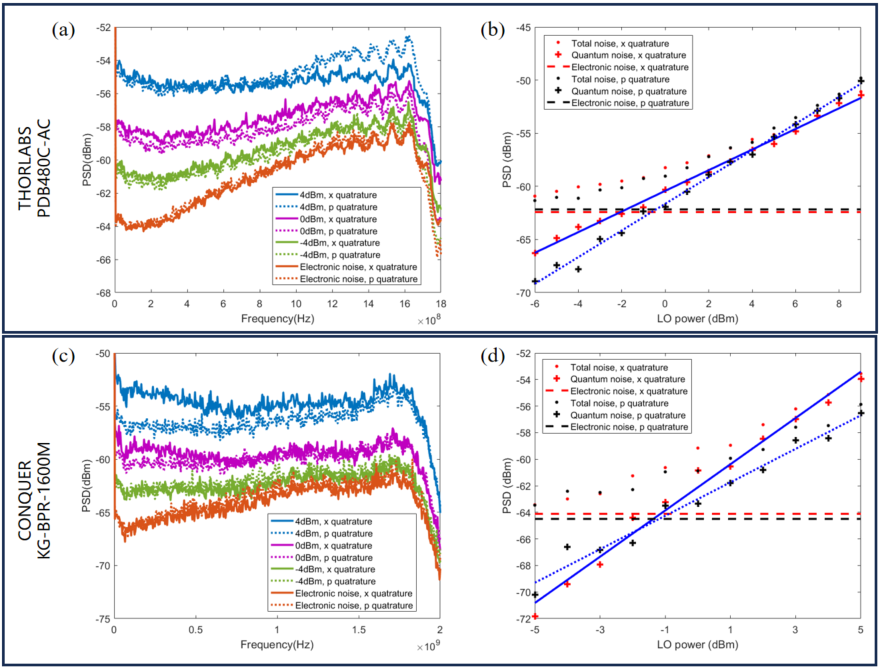}
\caption{\label{fig2} A laser provides the LO signal and connects to a variable optical attenuator, whose output signal is connected to the input port of the homodyne detector through a $50:50$ beam splitter, then the output port is connected to the spectrometer and the waveform is observed \cite{zhang20181}. (\textbf{a}) and (\textbf{c}) respectively present the average PSD of two pairs of homodyne detectors for different incident laser powers. (\textbf{b}) and (\textbf{d}) respectively present the electronic noise and total noise measured at 500MHz for different incident LO powers, with the quantum shot noise labeled. The blue curve represents the PSD fit for quantum shot noise. (\textbf{a}) and (\textbf{b}) for THORLABS PDB480C-AC, while (\textbf{c}) and (\textbf{d}) for CONQUER KG-BPR-1600M.}
\end{figure*}

In addition, to evaluate the asymmetry of the heterodyne detection, we measure the vacuum fluctuation power spectral density (PSD) of two pairs of PDB480C-AC \cite{480C} and KG-BPR-1600M \cite{kg} homodyne detectors respectively for different incident laser powers \cite{zhang20181} and conduct a comparative analysis. To be specific, a 1550 nm continuous laser is coupled to the homodyne detector via a single-mode fiber connected to the LO port, with the quantum signal port input in vacuum state. By varying the incident laser power, we can obtain different noise PSDs, as shown in Fig.~\ref{fig2} (\textbf{a}) and (\textbf{c}). It can be seen obviously that the PSD increases with the LO pump power, which is not completely coexisting at the same LO power. In addition, we also analyze the characteristics of the detector's electronic noise and the vacuum state shot noise. 
Noise PSD in (\textbf{a}) and (\textbf{c}) is composed of quantum shot noise and electronic noise. The electronic noise is mainly caused by the inherent electronic noise of photodiode and the amplification circuit, which is measured when the LO laser is turned off. We extract the average value of PSD at 500 MHz under different incident LO powers, and then plot the relationship between PSD and LO power, as shown in Fig.~\ref{fig2} (\textbf{b}) and (\textbf{d}). According to the PSD of total noise and electronic noise, the quantum shot noise is calculated, and the fitting curve of linear relationship with LO power is obtained. Comparing the measurement data of the two pairs of homodyne detectors, it is obvious that the electronic noise of the two detectors is not exactly the same, as well as the quantum shot noise collected by them, and the two fitted lines can not coincide at last. All the experiment results verify that there is asymmetry in the heterodyne detection, which will threaten the practical security of the protocol and cause the reduction of secret key rate.

\begin{table}[t]
\renewcommand\arraystretch{1.5}
\caption{\label{tab1}%
Measurement data in the case of quantum signal connected to the first input port of the beam splitter
}
\begin{ruledtabular}
\begin{tabular}{cccc}
\multirow{2}{*}{\textbf{Input1 (dBm)}} & \multicolumn{2}{c}{\textbf{Output (dBm)}} & \multirow{2}{*}{\textbf{Splitting ratio}} \\
                                       & \textbf{1}          & \textbf{2}          &                                           \\
                                       \colrule
-40                                    & -43.12              & -43.19              & 50.04:49.96                               \\
-35                                    & -38.19              & -38.46              & 50.18:49.82                               \\
-30                                    & -33.32              & -33.49              & 50.13:49.87                               \\
-25                                    & -28.24              & -28.37              & 50.11:49.89                               \\
-20                                    & -23.28              & -23.43              & 50.16:49.84                              
\end{tabular}
\end{ruledtabular}
\end{table}
\begin{table}[t]
\renewcommand\arraystretch{1.5}
\caption{\label{tab2}%
Measurement data in the case of local oscillator connected to the second input port of the beam splitter
}
\begin{ruledtabular}
\begin{tabular}{cccc}
\multirow{2}{*}{\textbf{Input2 (dBm)}} & \multicolumn{2}{c}{\textbf{Output (dBm)}} & \multirow{2}{*}{\textbf{Splitting ratio}} \\
                                       & \textbf{1}          & \textbf{2}          &                                           \\
                                       \colrule
6                                      & 2.57                & 2.76                & 48.22:51.78                           \\
7                                      & 3.59                & 3.74                & 48.98:51.02                           \\
8                                      & 4.57                & 4.75                & 49.03:50.97                           \\
9                                      & 5.56                & 5.73                & 49.25:50.75                           \\
10                                     & 6.59                & 6.74                & 49.44:50.56                          
\end{tabular}
\end{ruledtabular}
\end{table}

\begin{table}[htp]
\renewcommand\arraystretch{1.5}
\caption{\label{tab3}%
Electronic noise normalized by SNU for PDB480C-AC and KG-BPR-1600M, where $x$ and $p$ quadratures are calibrated separately
}
\begin{ruledtabular}
\begin{tabular}{cccc}
\multirow{2}{*}{\textbf{\begin{tabular}[c]{@{}c@{}}Homodyne\\ detector\end{tabular}}} & \multirow{2}{*}{\textbf{LO power(dBm)}} & \multicolumn{2}{c}{\textbf{$v_{el}$(SNU)}} \\
                                                                                      &                                         & \textbf{$x$}           & \textbf{$p$}           \\
                                                                                      \colrule
\multirow{3}{*}{PDB480C-AC}                                                           & 9                                       & 0.1403               & 0.0743               \\
                                                                                      & 8                                       & 0.1706               & 0.1009               \\
                                                                                      & 7                                       & 0.2061               & 0.1276               \\
                                                                                      \colrule
\multirow{3}{*}{KG-BPR-1600M}                                                              & 5                                       & 0.091                & 0.1545               \\
                                                                                      & 4                                       & 0.1327               & 0.1991               \\
                                                                                      & 3                                       & 0.1841               & 0.2897              
\end{tabular}
\end{ruledtabular}
\end{table}

\begin{figure*}[t]
\includegraphics[width=16cm]{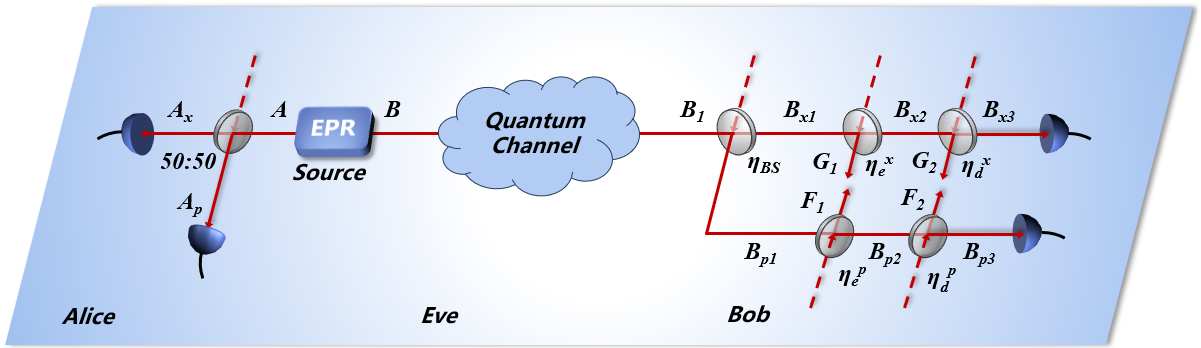}
\caption{\label{fig3} The entanglement-based scheme of the biased No-Switching CV-QKD protocol with a modified practical detector model, where two beam splitters are
applied to represent the electronic noise and the limited
detection efficiency. In this model, the heterodyne detector is composed of a beam spilitter with optimal transmittance $\eta_{BS}$ and two different homodyne detectors. We take the limited detection efficiency on Bob's side, which can be regarded as a trusted loss, while electronic noise is considered as channel loss. $\eta_{e}^x$ and $\eta_{e}^p$ describe the electronic noise of the two homodyne detectors respectively, while $\eta_{d}^x$ and $\eta_{d}^p$ describe the detection efficiency of the detectors.}
\end{figure*}

Finally, according to the spectrum diagam and data in Fig.~\ref{fig2}, we can calculate the electronic noise $v_{el}$ normalized by SNU of the two kinds of homodyne detectors under the one-time calibration model, where the SNU in this model is measured directly as the output when the LO signal is on, which is equivalent to total noise.  The power of total noise and electronic noise at 1 GHz is extracted and derived respectively, so $v_{el}$ can be obtained by
\begin{equation}
v_{el}=10^{(P_{el}-P_{SNU})/10},
\end{equation}
where both $P_{el}$ and $P_{SNU}$ can be directly extracted from the PSD diagram. The calculation results are shown in Table~\ref{tab3}, presenting the of presenting the asymmetry of electronic noise.

\section{\label{sec:level3}Practical Biased No-Switching System}
In this section, a modified biased No-Switching protocol is proposed for the practical systems, which can better match the practical PM model and close the security loophole. An optimized secret key rate is achieved by further adjusting the transmittance of the beam splitter. The security and feasibility of the biased protocol is analyzed, and the detailed calculation of the secret key rate in the asymptotic-limit and finite-size regime is given.
\subsection{The Biased No-Switching Protocol}
In this protocol, we take into account the imperfections of beam splitter and the differences in homodyne detectors, where the transmittance of the beam splitter is defined as a more generalized parameter and the two probes of homodyne detection are calibrated respectively to obtain different detector parameters for calculation. This approach effectively eliminates the practical security loophole in practical systems. As for the reduction of secret key rate caused by asymmetry in the biased protocol, we investigate the relationship between the secret key rate and the transmittance of the beam splitter at different distances. It can be found that the transmittance of the beam splitter, denoted as $\eta_{BS}$, deviates from 50\% for the optimal key rate. Therefore, correcting the transmittance of the beam splitter to the optimal value corresponding to different distances can improve the key rate and optimize the performance of the protocol.

The EB model of biased No-Switching protocol with a modified detector model is shown as Fig.~\ref{fig3}, where the improvement is mainly carried out at the receiving end. The imperfections of two homodyne detectors are separately calibrated after the first beam splitter with the optimal transmittance. In the $x$ quadrature, the electronic noise of the detector is described by a beam splitter with transmittance $\eta_e^x$, while the detection efficiency is described by another beam splitter with transmittance $\eta_d^x$. Similarly, in the $p$ quadrature, the electronic noise and detection efficiency are described by different beam splitters with transmittance $\eta_e^p$ and $\eta_d^p$ respectively. In this modified detector model, the electronic noise no longer need to be calibrated in the process of modeling the detection efficiency, which realizes one-time calibration of the detector. Therefore, this model is not affected by environmental factors such as temperature and time on electronic noise, which can greatly improve the flexibility of actual detector modeling, reduce the required calibration links and complex calibration processes in the system, and improve the accuracy of detector modeling, so it is closer to the actual commercial implementation, and the subsequent analysis in this study will primarily concentrate on examining and evaluating this modified model.


\subsection{Secret Key Rate Calculation}
As for the modified protocol, the secret key rate is calculated in the same way as in Eq.~\eqref{(4)}, while the mutual information between Alice and Bob can be written as 
\begin{equation}
    \begin{aligned}
	    I_{AB}&=\frac{1}{2}\log_2\frac{V_{A_x}}{V_{A_x|B_x}}+\frac{1}{2}\log_2\frac{V_{A_p}}{V_{A_p|B_p}} \\
        &=\frac{1}{2}\log_2\frac{V+\chi_{tot}^x}{1+\chi_{tot}^x}+\frac{1}{2}\log_2\frac{V+\chi_{tot}^p}{1+\chi_{tot}^p}
	\end{aligned}
\end{equation}
	
where $\chi_{tot}^x=(\chi_{line}+\chi_{hom}^x)/T$ and $\chi_{tot}^p=(\chi_{line}+\chi_{hom}^p)/T$, and the detection-added noise referred to Bob’s side is respectively defined as $\chi_{hom}^x=(1-\eta_d^x\eta_{BS}+v_{el}^x)/(\eta_d^x\eta_{BS})$ and $\chi_{hom}^p=(1-\eta_d^p\eta_{BS}+v_{el}^p)/(\eta_d^p\eta_{BS})$.

The mutual information between Eve and Bob is given by the Holevo bound of their von Neumann entropy $\chi_{BE}$,
\begin{equation}
	\begin{aligned}
		\chi_{BE}&=S(\rho_{A_xA_pG_2F_2B_{x_3}B_{p_3}})-S(\rho_{A_xA_pG_2F_2}^{p_B,x_B})  \\
  &=\sum_{i=1}^6G(\frac{\lambda_i-1}{2})-\sum_{i=7}^{10}G(\frac{\lambda_i-1}{2}),
	\end{aligned}
\end{equation}
here $\lambda_{1\sim6}$ is the symplectic eigenvalue of the covariance matrix $\gamma_{A_xA_pG_2F_2B_{x_3}B_{p_3}}$ and $\lambda_{7\sim10}$ is the symplectic eigenvalue of the covariance matrix $\gamma_{A_xA_pG_2F_2}^{p_B,x_B}$.

$\gamma_{A_xA_pG_2F_2}^{p_B,x_B}$ is the covariance matrix obtained by Bob through heterodyne detection. It can be derived from matrix $\gamma_{A_xA_pG_2F_2B_{x_3}B_{p_3}}$ by extracting the mode $p$ and mode $x$ separately. After extracting the mode $p$, it becomes
\begin{multline}
\gamma_{A_xA_pG_2F_2B_{x_3}}^{p_B}=\gamma_{A_xA_pG_2F_2}- \\
\sigma_{A_xA_pG_2F_2B_{x_3}B_{p_3}}^TH_{hom}^p\sigma_{A_xA_pG_2F_2B_{x_3}B_{p_3}},
\end{multline}
and $H_{hom}^p=(X\gamma_{B_{p_3}}X)^{MP}$, $X=
\begin{bmatrix}
		1 & 0 \\
		0 & 0
\end{bmatrix}$, the other three matrices are given by the covariance matrix:
\begin{multline}
\gamma_{A_xA_pG_2F_3B_{x_3}B_{p_3}}= \\
\begin{bmatrix}
\gamma_{A_xA_pG_2F_2B_{x_3}} & \sigma_{A_xA_pG_2F_2B_{x_3}B_{p_3}}^T \\
\sigma_{A_xA_pG_2F_2B_{x_3}B_{p_3}} & \gamma_{B_{p_3}}
\end{bmatrix},
\end{multline}
next we extract the mode $x$ and get that
\begin{multline}
\gamma_{A_xA_pG_2F_2}^{p_B,x_B}=\gamma_{A_xA_pG_2F_2}- \\
\sigma_{A_xA_pG_2F_2B_{x_3}}^TH_{hom}^x\sigma_{A_xA_pG_2F_2B_{x_3}},
\end{multline}
here $H_{hom}^x=(P\gamma_{B_{x_3}}P)^{MP}$, $P=
\begin{bmatrix}
		0 & 0 \\
		0 & 1
\end{bmatrix}$, the other three matrices are given by the covariance matrix:
\begin{multline}
\gamma_{A_xA_pG_2F_2B_{x_3}}^{p_B}= \\
\begin{bmatrix}
\gamma_{A_xA_pG_2F_2} & \sigma_{A_xA_pG_2F_2B_{x_3}}^T \\
\sigma_{A_xA_pG_2F_2B_{x_3}} & \gamma_{B_{x_3}}
\end{bmatrix},
\end{multline}
therefore, our ultimate goal is to obtain the matrix $\gamma_{A_xA_pG_2F_2B_{x_3}B_{p_3}}$, so we need to model the beam splitter with transmittance $\eta_{BS}$, and the two heterodyne detectors separately.

The mode A from Alice passes through the quantum channel and gets the covariance matrix $\gamma_{AB_1}$. $\gamma_{AB_1}$ only depends on the system composed of Alice and quantum channel, and can be written as
\begin{equation}
	\begin{aligned}
\gamma_{AB_1}&=
\begin{bmatrix}
\gamma_A & \sigma_{AB_1}^T \\
\sigma_{AB_1} & \gamma_{B_1}
\end{bmatrix} \\
&= 
\begin{bmatrix}
			VI_2 & \sqrt{T(V^2-1)}\sigma_z \\
			\sqrt{T(V^2-1)}\sigma_z & T(V+\chi_{line})I_2
\end{bmatrix}, \label{16}
	\end{aligned}
\end{equation}
where $V$ is the variance of the EPR state in the EB model and $V=V_A+1$, $I_2$ is the $2\times2$ identity matrix, $\sigma_z=
	\begin{bmatrix}
		1 & 0 \\
		0 & -1
	\end{bmatrix}$.
 
After the first beam splitter, mode A and mode B will be divided into two modes to get the covariance matrix $\gamma_{A_xA_pB_{p_1}B_{x_1}}$, after mode A is divided, we can get
	\begin{equation}
		\gamma_{A_xA_pB_1}=(Y^{BS}_A)(I_2\oplus\gamma_{AB_1})(Y^{BS}_A)^T,
	\end{equation}
$Y^{BS}_A$ describes the transmittance of the beam splitter for mode A, which is used to model the detection efficiency of the detector,
	\begin{equation}
		Y^{BS}_A=(Y_{A_xA_p}^{BS})\oplus I_2,
	\end{equation}
	\begin{equation}
		Y_{A_xA_p}^{BS}=
		\begin{bmatrix}
			\sqrt{\eta}I_2 & \sqrt{1-\eta}I_2 \\
			-\sqrt{1-\eta}I_2 & \sqrt{\eta}I_2
		\end{bmatrix},
	\end{equation}
where $\eta$=0.5, thus we can further split the mode B to obtain
	\begin{equation}
		\gamma_{A_xA_pB_{p_1}B_{x_1}}=(Y^{BS}_{B_1})(\gamma_{A_xA_pB_1}\oplus I_2)(Y^{BS}_{B_1})^T,
	\end{equation}
$Y^{BS}_{B_1}$ describes the transmittance of the beam splitter for mode $B_1$,
	\begin{equation}
		Y^{BS}_{B_1}=I_2\oplus I_2\oplus(Y_{B_xB_p}^{BS}),
	\end{equation}
	\begin{equation}
		Y_{B_xB_p}^{BS}=
		\begin{bmatrix}
			\sqrt{\eta_{BS}}I_2 & \sqrt{1-\eta_{BS}}I_2 \\
			-\sqrt{1-\eta_{BS}}I_2 & \sqrt{\eta_{BS}}I_2
		\end{bmatrix}.
	\end{equation}
 Then, mode $x$ and mode $p$ are respectively detected, and finally the covariance matrix $\gamma_{A_xA_pG_2F_2B_{x_3}B_{p_3}}$ is obtained. While detecting the x mode, $Y^{BS}_{x_1}$ models the electronic noise of the detector,
	\begin{equation}
		Y^{BS}_{x_1}=I_2\oplus I_2\oplus I_2\oplus(Y_{\eta_{e}^x}^{BS}),
	\end{equation}
	\begin{equation}
		Y_{\eta_{e}^x}^{BS}=
		\begin{bmatrix}
			\sqrt{\eta_{e}^x}I_2 & \sqrt{1-\eta_{e}^x}I_2 \\
			-\sqrt{1-\eta_{e}^x}I_2 & \sqrt{\eta_{e}^x}I_2
		\end{bmatrix}.
	\end{equation}
then we can get 
         \begin{multline}
        \gamma_{A_xA_pB_{p_1}B_{x_2}G_1}= \\
        (Y^{BS}_{x_1})(\gamma_{A_xA_pB_{p_1}B_{x_1}} \oplus I_2)(Y^{BS}_{x_1})^T,
        \end{multline}
$Y^{BS}_{x_2}$ is used to model the detection efficiency of the detector,
	\begin{equation}
		Y^{BS}_{x_2}=I_2\oplus I_2\oplus I_2\oplus Y_{\eta_{d}^x}^{BS},
	\end{equation}
	\begin{equation}
		Y_{\eta_{d}^x}^{BS}=
		\begin{bmatrix}
			\sqrt{\eta_{d}^x}I_2 & \sqrt{1-\eta_{d}^x}I_2 \\
			-\sqrt{1-\eta_{d}^x}I_2 & \sqrt{\eta_{d}^x}I_2
		\end{bmatrix},
	\end{equation}
the next covariance matrix is given by
         \begin{multline}
        \gamma_{A_xA_pB_{p_1}B_{x_3}G_2}= \\
        (Y^{BS}_{x_2})(\gamma_{A_xA_pB_{p_1}B_{x_2}} \oplus I_2)(Y^{BS}_{x_2})^T.
        \end{multline}
The matrice $\gamma_{A_xA_pB_{p_1}B_{x_2}}$ can be derived from the decomposition of the covariance matrix $\gamma_{A_xA_pB_{p_1}B_{x_2}G_1}$. 

Then the $p$ mode is detected, $Y^{BS}_{p_1}$ models the electronic noise of the detector,
	\begin{equation}
		Y^{BS}_{p_1}=I_2\oplus I_2\oplus I_2\oplus I_2\oplus Y_{\eta_{e}^p}^{BS},
	\end{equation}
	\begin{equation}
		Y_{\eta_{e}^p}^{BS}=
		\begin{bmatrix}
			\sqrt{\eta_{e}^p}I_2 & \sqrt{1-\eta_{e}^p}I_2 \\
			-\sqrt{1-\eta_{e}^p}I_2 & \sqrt{\eta_{e}^p}I_2
		\end{bmatrix},
	\end{equation}
then we can get 
         \begin{multline}
        \gamma_{A_xA_pB_{x_3}G_2B_{p_2}F_1}= \\
        (Y^{BS}_{p_1})(\gamma_{A_xA_pB_{x_3}G_2B_{p_1}}\oplus I_2)(Y^{BS}_{p_1})^T,
        \end{multline}
$\gamma_{A_xA_pB_{x_3}G_2B_{p_1}}$ can be derived with appropriate rear rangement of lines and columns from the matrixi $\gamma_{A_xA_pB_{p_1}B_{x_3}G_2}$. $Y^{BS}_{p_2}$ is used to model the detection efficiency of the detector,
	\begin{equation}
		Y^{BS}_{p_2}=I_2\oplus I_2\oplus I_2\oplus I_2\oplus Y_{\eta_{d}^p}^{BS},
	\end{equation}
	\begin{equation}
		Y_{\eta_{d}^p}^{BS}=
		\begin{bmatrix}
			\sqrt{\eta_{d}^p}I_2 & \sqrt{1-\eta_{d}^p}I_2 \\
			-\sqrt{1-\eta_{d}^p}I_2 & \sqrt{\eta_{d}^p}I_2
		\end{bmatrix},
	\end{equation}
so the final covariance matrix is given by
         \begin{multline}
        \gamma_{A_xA_pB_{x_3}G_2B_{p_3}F_2}= \\
        (Y^{BS}_{p_2})(\gamma_{A_xA_pB_{x_3}G_2B_{p_2}} \oplus I_2)(Y^{BS}_{p_2})^T,
        \end{multline}
here, the matrix $\gamma_{A_xA_pB_{x_3}G_2B_{p_2}}$ can be derived from the decomposition of the covariance matrix $\gamma_{A_xA_pB_{x_3}G_2B_{p_2}F_1}$.

Similarly, $\gamma_{A_xA_pG_2F_2B_{x_3}B_{p_3}}$ is obtained by matrix $\gamma_{A_xA_pB_{x_3}G_2B_{p_3}F_2}$ through columns and rows transformation, so that we have the required elements for calculation on $\gamma_{A_xA_pG_2F_2B_{x_3}B_{p_3}}$ and $\gamma_{A_xA_pG_2F_2}^{p_B,x_B}$, and then we can find the symplectic eigenvalues of the matrices.
\subsection{Finite-Size Analysis}
In this section, we introduce the influence of finite-size statistical fluctuation on the security of the biased No-Switching protocol, and mainly analyze its effect on the parameter estimation process, more precisely, the effect of finite-size block on excess noise $\varepsilon$ and channel transmissivity $T$. We study the parameter estimation procedure without post-selection, and finally give the covariance matrix used to calculate the secret key, which is suitable for both the conventional detector model and the modified detector model.

The statistical fluctuation of the sampling estimation will become worse on account of the finite-size block, which will lead to a decrease in the accuracy of the evaluation of Eve's eavesdropping behavior in the channel. Therefore, we need to widen the fluctuation space of the estimate to include more possible cases, and this relaxed estimation leads to the deterioration of the protocol performance. The significant impact of finite-size statistical fluctuation on CV-QKD system has been widely demonstrated \cite{leverrier2010finite}, and the revised secret key rate can be expressed as:
\begin{equation}
		R=\frac{n}{N}[\beta I(A:B)-S_{\epsilon_{_{PE}}}(B:E)-\Delta (n) ],
\end{equation}
here $N$ is the total length of the data exchanged between Alice and Bob, $n$ of which is used to generate the key, and the remaining number $m=N-n$ of the samples is used for parameter evaluation. The parameter $\Delta(n)$ is related to the security of private key amplification and its value is expressed as:
\begin{equation}
		\Delta(n)=(2dimH_X+3)\sqrt{\frac{log_2(2/\bar{\epsilon})}{n}}+\frac{2}{n}log_2(1/\epsilon_{PA}),
\end{equation}
where $H_X$ is the Hilbert space corresponding to the variable from Alice used in the raw key. $\bar{\epsilon}$ is a smoothing parameter, and $\epsilon_{PA}$ is the failure probability of the privacy key amplification procedure. Both of them are intermediate variables that can be specified arbitrarily and their values are as small as possible.

Considering the effect of finite-size block on excess noise and channel transmissivity estimation, we express the usual conditional entropy $S(B:E)$ as $S_{\epsilon_{_{PE}}}(B:E)$, which is the maximum value of Holevo information between Eve and Bob in the case of statistical fluctuations in parameter estimation. In the absence of post-selection, the upper bound can be calculated using only the entanglement-based covariance matrix shared between Alice and Bob. The original covariance matrix $\gamma_{AB_1}$ is the same as Eq.~(\ref{16}).

The covariance matrix under the condition of finite-size regime is given by $\gamma_{\epsilon_{_{PE}}}$, and its estimate is obtained through the samples of $m$ related variables$(x_i,y_i)_{1 \ldots m}$. $x$ and $y$ are the classical data of Alice and Bob after quantum state measurement respectively. For a general linear channel, the relationship between Alice and Bob's data is
\begin{equation}
        y=tx+z,
\end{equation}
through the analysis of the variance of the data from Bob, it can be known that $t=\sqrt{T}$, $z$ follows a central normal distribution with the shape of $\sigma^2=1+T\varepsilon$, and $x$ is a random variable obeying the normal distribution of the variance $V_A$ for Gaussian modulation. So our goal is to study the relationship between $S(B:E)$ and the variables $t$ and $\sigma^2$. For any
value of the modulation variance, The following inequalities are established:
\begin{equation}
        \frac{\partial S(B:E)}{\partial t}\bigg|_{\sigma^2}<0, \frac{\partial S(B:E)}{\partial \sigma^2}\bigg|_{t}>0,
\end{equation}
this means a covariance matrix $\gamma_{\epsilon_{_{PE}}}$ that minimizes the secret key rate can be found with the probability at least $1-\epsilon_{PE}$,
	\begin{equation}
\gamma_{\epsilon_{_{PE}}}=
\begin{bmatrix}
			VI_2 & t_{\min}Z\sigma_z \\
			t_{\min}\sqrt{(V^2-1)}\sigma_z & (t_{\min}(V-1)+\sigma^2_{\max})I_2
\end{bmatrix}, 
	\end{equation}
where $t_{\min}$ and $\sigma^2_{\max}$ calculated from the sample data are respectively the minimum value of $t$ and the maximum value of $\sigma^2$ when the error probability is $\epsilon_{_{PE}}/2$. Maximum-likelihood estimators $\hat{t}$ and $\hat{\sigma}^2$ are known for the normal linear model:
\begin{equation}
        \hat{t}=\frac{\sum_{i=1}^m x_iy_i}{\sum_{i=1}^m x_i^2}, \hat{\sigma}^2=\frac{1}{m}\sum_{i=1}^m(y_i-\hat{t}x_i)^2.
\end{equation}
Using the theorem of large numbers and the central limit theorem, it can be seen that $\hat{t}$ and $\hat{\sigma}^2$ approximate the following distribution:
\begin{equation}
        \hat{t}\sim \mathcal{N}(t,\frac{\sigma^2}{\sum_{i=1}^m x_i^2}), \frac{m\hat{\sigma}^2}{\sigma^2}\sim \chi^2(m-1),
\end{equation}
where $\mathcal{N}$ represents normal distribution, $\chi^2$ represents chi-square distribution, $t$ and $\sigma^2$ are the true values of the parameters, so it is possible to calculate the lower bound $t_{\min}$ of the confidence interval for $t$ and the upper bound $\sigma^2_{\max}$ for $\sigma^2$ when the confidence probability is $\epsilon_{_{PE}}/2$.
\begin{equation*}
        t_{\min}\approx\hat{t}-z_{\epsilon_{_{PE}}/2}\sqrt{\frac{\hat{\sigma}^2}{m(V-1)}}
\end{equation*}
\begin{equation} \sigma^2_{\max}\approx\hat{\sigma}^2+z_{\epsilon_{_{PE}}/2}\frac{\hat{\sigma}^2\sqrt{2}}{\sqrt{m}},
\end{equation}
where $z_{\epsilon_{_{PE}}/2}$ satisfies that $1-\mathrm{erf}(z_{\epsilon_{_{PE}}/2}/\sqrt{2})=\epsilon_{PE}/2$ and $\mathrm{erf}$ is the error function which defined as
\begin{equation} 
\mathrm{erf}(x)=\frac{2}{\sqrt{\pi}}\int_0^xe^{-t^2}dt.
\end{equation}

To continue the analysis of the security of the protocol theoretically, we take the expected values of $\hat{t}$ and $\hat{\sigma}^2$ ($E[\hat{t}]=\sqrt{T}$, $E[\hat{\sigma}^2]=1+T\varepsilon$) instead of the maximum likelihood estimation, then $t_{\min}$ and $\sigma^2{\max}$ can be calculated as
\begin{equation*}
        t_{\min}\approx\sqrt{T}-z_{\epsilon_{_{PE}}/2}\sqrt{\frac{\hat{\sigma}^2}{m(V-1)}}
\end{equation*}
\begin{equation} \sigma^2_{\max}\approx1+T\varepsilon+z_{\epsilon_{_{PE}}/2}\frac{\hat{\sigma}^2\sqrt{2}}{\sqrt{m}}.
\end{equation}
Finally, we can obtain the covariance matrix $\gamma_{\epsilon_{_{PE}}}$, 
\begin{equation}
\gamma_{\epsilon_{_{PE}}}=\gamma_{AB_1}+
\begin{bmatrix}
			0 & \Delta_Z\sigma_z \\
			\Delta_Z\sigma_z & \Delta_B
\end{bmatrix}, 
	\end{equation}
where
\begin{equation*}
        \Delta_Z=-z_{\epsilon_{_{PE}}/2}\sqrt{\frac{1+T\varepsilon}{m(V-1)}}
\end{equation*}
\begin{multline} \Delta_B=\frac{z_{\epsilon_{_{PE}}/2}}{\sqrt{m}}((1+T\varepsilon)\sqrt{2}-2\sqrt{T(V-1)})\\
+z_{\epsilon_{_{PE}}/2}^2\frac{1+T\varepsilon}{m}.
\end{multline}

\begin{figure*}[htp]
\includegraphics[width=17cm]{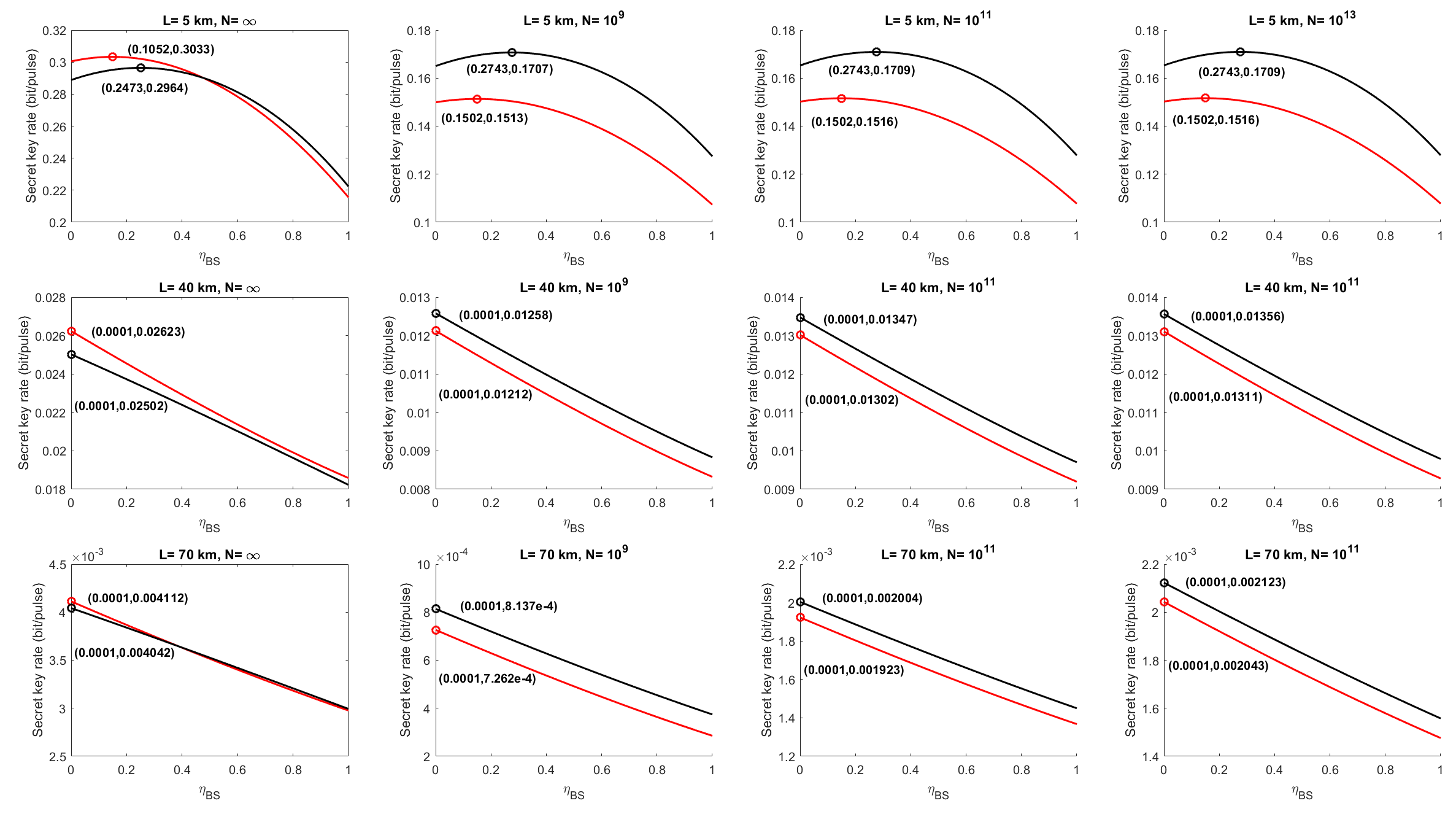}
\caption{\label{fig4} The relationship between secret key rate and $\eta_{BS}$ against the conventional model described by the black line and the modified model described by the red line, with the optimal $\eta_{BS}$ marked for different distances. Each row reflects the optimal $\eta_{BS}$ at L = 10 km, 50 km and 70 km respectively, and each column represents the relationship between the secret key rate and $\eta_{BS}$ at different values of N, where the first column is for the asymptotic-limit regime, while the second, third, and fourth columns are for different cases when $N=10^9$, $10^{11}$ and $10^{13}$ respectively. The reconciliation efficiency $\beta$ = 95\%, excess noise $\varepsilon$ = 0.05, the variance $V$ = 5.}
\end{figure*}

The final covariance matrix can be used to calculate the secret key rate under the finite-size regime, taking into account the influence of statistical fluctuations in parameter estimation. Before proceeding to the simulation experiments in the next section, we first give a hint of the block length and security parameters that are generally required at a given distance. For long-distance CV-QKD, the main effect is the uncertainty on excess noise, which can be approximated as
\begin{equation} 
\label{47}
\Delta_m\varepsilon \approx \frac{z_{\epsilon_{_{PE}}/2}\sqrt{2}}{T\sqrt{m}}.
\end{equation}
According to the above Eq.~(\ref{47}) we can obtain
\begin{equation} 
m \approx \frac{2z_{\epsilon_{_{PE}}/2}^2}{T^2\Delta_m\varepsilon^2}.
\end{equation}
For $\epsilon_{PE}=10^{-10}$, then $z_{\epsilon_{_{PE}}/2}=6.5$, and if $\Delta_m\varepsilon$=1/100 is required, then the relationship between the sample size required for parameter estimation and the channel transmissivity is approximately
\begin{equation} 
m \propto \frac{10^6}{T^2}.
\end{equation}
If the distance between Alice and Bob is 50 km, then $T=10^{-1}$, which means that the block length should be of the order of $10^8$. If the distance is 100 km, then the block length should be of the order of $10^9$.

\section{\label{sec:level4}Simulation results and discussion}
In this section, we provide some simulation results and analyses to make a comparison on the ideal No-Switching protocol and the biased No-Switching protocol with both the conventional detector model and the modified detector model, 
as well as the results of modified model in the finite-size regime. The following simulation experiments are based on reverse reconciliation under the collective attack, and the parameters for the heterodyne detector model are ($\eta_x$, $\eta_p$, $v_{el}^x$, $v_{el}^p$) = (60\%, 80\%, 0.1403, 0.0743), where the electronic noise of the two quadratures is based on the experiment results in Sec.~\ref{sec:level2}.

Firstly, we compare the performance of the ideal No-Switching protocol and the biased No-Switching protocol, where the secret key rate of the ideal protocol is always higher than that of the biased protocol. The simulation results are given together with the optimized rates in the following analysis. So it is proved that there is definitely a practical security loophole caused by the mismatch between the EB model and the PM model, which indicates the great importance to consider the asymmetry of heterodyne detection in practical system.

\begin{figure*}[t]
\centering
\includegraphics[width=17cm]{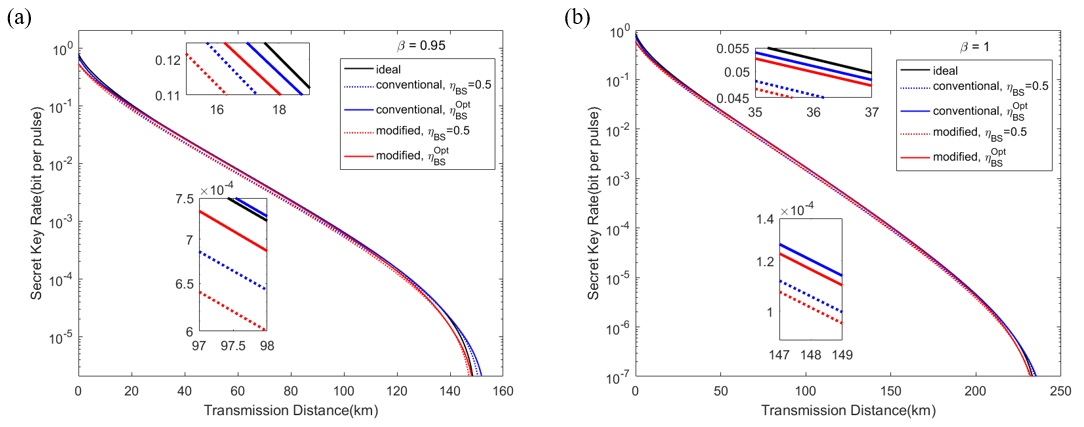}
\caption{ The comparison diagram of the secret key rate between the ideal No-Switching protocol and the biased No-Switching protocol with both conventional detector model and the modified detector model, as well as the key rate after correction on $\eta_{BS}$. The black solid line describes the key rate-distance relationship for the ideal No-Switching model. The blue dotted line describes the key rate-distance relationship for the conventional model when $\eta_{BS}$ = 0.5, while the blue solid one describes the key rate-distance relationship for the conventional model when $\eta_{BS}$ is given the optimal value. The red dotted line describes the key rate-distance relationship for the modified model when $\eta_{BS}$ = 0.5, while the red solid one describes the key rate-distance relationship for the modified model when $\eta_{BS}$ is given the optimal value. The excess noise $\varepsilon$ = 0.05, the variance $V$ = 5. For figure (\textbf{a}), the value of the reconciliation efficiency $\beta$ is 95\% while in figure (\textbf{b}), $\beta$ = 1. \label{fig5}}
\end{figure*}

\begin{figure*}[t]
\centering
\includegraphics[width=17cm]{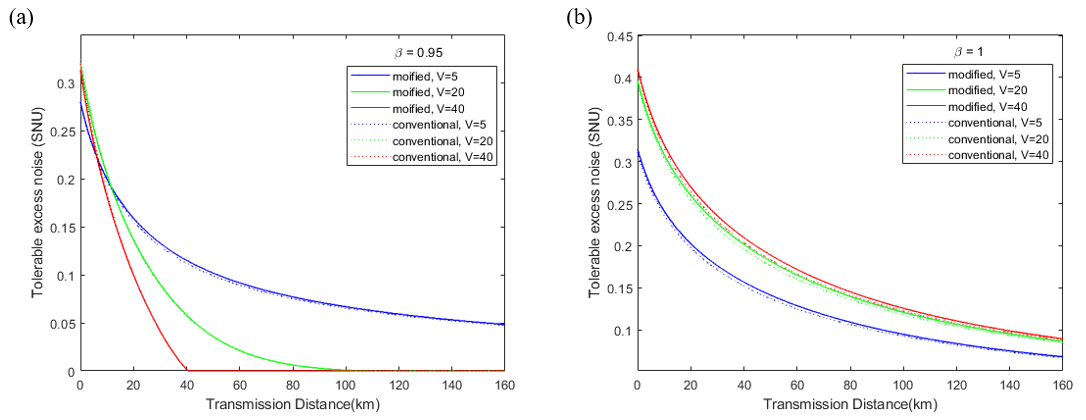}
\caption{ The maximal tolerable excess noise as a function of transmission distances for both conventional model and modified model which are described by dotted and solid lines respectively. (\textbf{a}) presents the tolerable excess noise of two models for different modulation variances $V=$ 5, 20 and 40, described by the blue lines, green lines and red lines respectively in the case of the reconciliation efficiency $\beta = 0.95$ while (\textbf{b}) presents that when $\beta = 1$. \label{fig6}}
\end{figure*}

Due to the asymmetry of the heterodyne detector in practice, the optimal calculation results of the secret key rate are always obtained when $\eta_{BS}$ deviates from 0.5, which is presented in Fig.~\ref{fig4}. We provide the relationship between the secret key rate and $\eta_{BS}$ for different transmission distances and block lengths, where the red lines represent the modified model and the black lines represent the conventional model. It can be seen that the optimal $\eta_{BS}$ values of the two models are approximately the same at the same distance, the asymmetry of the detector is smaller in the short distance transmission but becomes more significant for longer distances. In the finite-size regime, the difference in the optimal secret key rate between the two models is more obvious compared to the case of asymptotic-limit regime. At the same transmission distance, with the decrease of the block length, the secret key rates of the two models have a certain decline, but there is almost no difference for the optimal value of $\eta_{BS}$. On the whole, the detector asymmetry of the two models causes roughly the same bias at different distances and different block lengths.

Figure ~\ref{fig5} shows the comparison between the ideal No-Switching protocol and the biased No-Switching protocol with both conventional detector model and the modified detector model, presenting the original key rate and the modified rate after the $\eta_{BS}$ corrected to the optimal value. Different cases when $\beta=0.95$ and $\beta=1$ are also given. It can be seen that the secret key rate of ideal protocol has a slight uplift, which threatens the practical security of the system. Moreover, the key rate between the two detector model is almost identical, with the conventional model yielding slightly higher key rate. After optimizing the value of $\eta_{BS}$, the modified key rates are superior for any distance compared to that before correction. Whatever the value of $\beta$, the secret key rate of the conventional model is significantly higher than that of the modified model when the distance is more longer in both of the two figures. Therefore, the simplification of practical detector modeling may always come at the cost of sacrificing key rate, a balance can be found between practical modeling and the secret key rate by considering the overall performance of the system.

\begin{figure}[t]
\includegraphics[width=8.6cm]{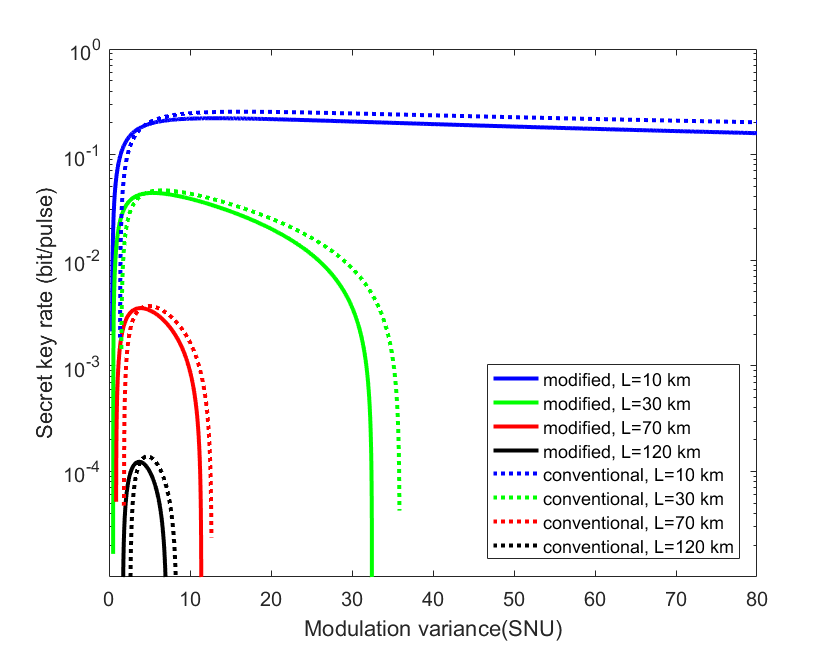}
\caption{The functional relationship between the secret key rate and modulation variance for the two different models. The four solid lines represent this relationship of the conventional detector model at different transmission distances, while the dashed lines reflect the modified detector model. We simulate at the cases when $L$ = 10 km, 30 km, 70 km and 120 km respectively.\label{fig7}}
\end{figure}  

\begin{figure}[bp]
\includegraphics[width=8.6cm]{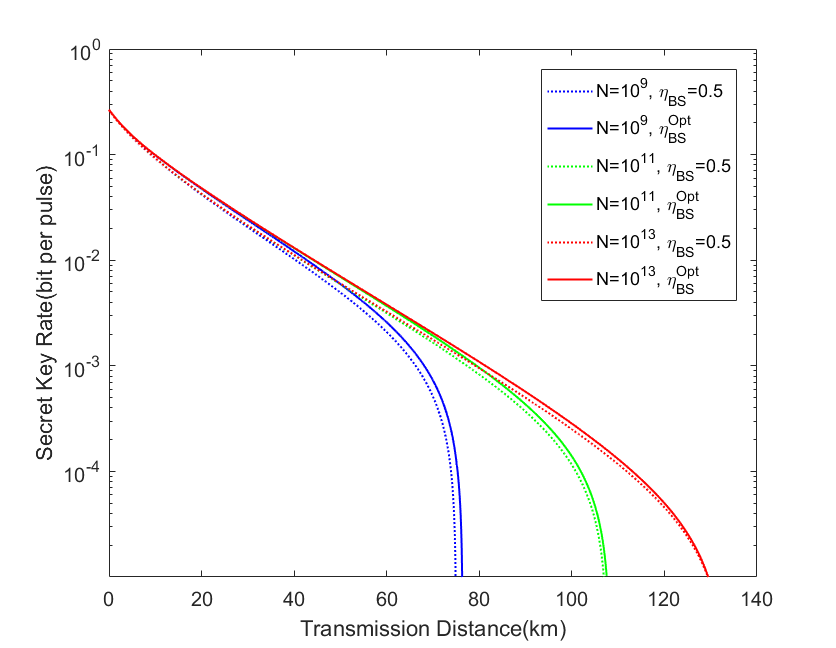}
\caption{ Both of the original and optimal secret key rate as a function of transmission distance for the modified model under different block lengths. The green, carmine and red lines show results for the block length $N$ of $10^{9}$, $10^{11}$ and $10^{13}$ respectively. The solid and dotted lines correspond to the original and optimal secret key rate respectively. The variance $V=5$, and the reconciliation efficiency $\beta = 95\%$. \label{fig8}}
\end{figure}  

\begin{figure}[t]
\includegraphics[width=8.6cm]{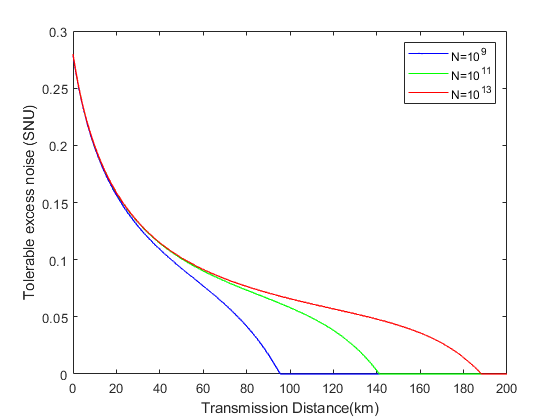}
\caption{ The maximal tolerable excess noise as a function of transmission distance for the modified model in the finite-size regime. The tolerable excess noise of the modified model for different block lengths $N=$ $10^{9}$, $10^{11}$ and $10^{13}$, described by the blue, green and red lines respectively in the case of the variance $V=5$ and the reconciliation efficiency $\beta = 95\%$. \label{fig9}}
\end{figure}  

The simulate results of maximal tolerable excess noise for the two models as a function of transmission distances with different modulation variances are shown in Fig.~\ref{fig6}, where the solid and dotted lines represent the modified model and the conventional model respectively. The tolerable excess noise is defined as the maximum excess noise that maintains the positive secret key rate at a fixed transmission distance. Fig.~\ref{fig6} (\textbf{a}) and (\textbf{b}) represent the maximal tolerable excess noise for a reconciliation efficiency of $\beta=0.95$ and the ideal scenario, respectively. It can be seen that with the increase of variance, the value of the tolerable excess noise decreases at a faster rate with the distance. Both models exhibit similar tolerance to excess noise at $V=$ 20 and 40, while at $V=5$, the tolerable noise of the modified model is slightly higher than that of the conventional model. For the ideal case of $\beta=1$, it is evident that the tolerable excess noise is significantly improved for different modulation variances, and for larger modulation variances, the tolerable noise decline rate will decrease significantly as the distance increases, indicating an improved tolerance of the protocol to excess noise.

Figure ~\ref{fig7} illustrates the impact of modulation variance on the secret key rate. It can be observed that the key rate initially increases with the increase in modulation variance. During this process, the key rate in the modified model is higher than that in the conventional detection model for different transmission distances. After obtaining the optimal modulation variance, the key rate decreases with further increase in modulation variance. Additionally, the rate of decrease in the key rate is higher in the modified model compared to the conventional model. From the simulation results, it is also apparent that the optimal modulation variance increases with the increase in distance. Therefore, in practical applications, the most suitable model and modulation variance can be selected to maximize the key rate based on these relationships.

\begin{figure}[t]
\includegraphics[width=8.6cm]{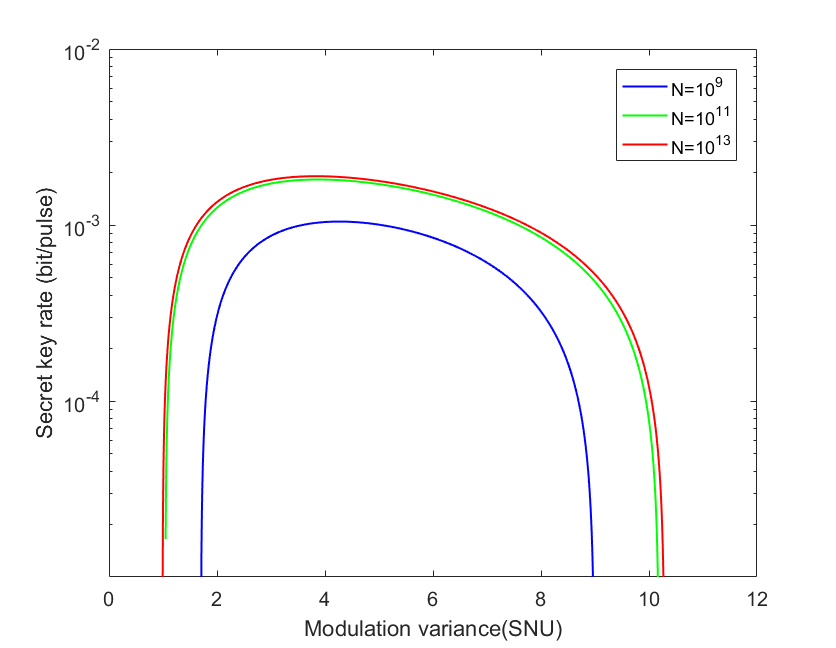}
\caption{ The secret key rate as a function of modulation variance for the modified model in the finite-size regime for different block lengths $N=$ $10^{9}$, $10^{11}$ and $10^{13}$, described by the blue, green and red lines respectively in the case of the transmission distance $L=80$ and the reconciliation efficiency $\beta = 95\%$. \label{fig10}}
\end{figure}

In addition, Fig.~\ref{fig8}, Fig.~\ref{fig9} and Fig.~\ref{fig10} show the performance of the biased No-Switching protocol with the modified detector model in the finite-size regime. Some of the parameters used in the simulation are provided as follows: the smoothing parameter $\bar{\epsilon}$, the failure probability of parameter estimation $\epsilon_{PE}$, the failure probability of the privacy amplification procedure $\epsilon_{PA}$ are all set to $10^{-10}$ \cite{leverrier2010finite}. The dimension of the Hilbert space of the variable in the raw key is set to $dimH_X = 2$.

Figure~\ref{fig8} shows the influence of block length $N$ to the secret key rate for the modified detector model and we provide the comparison of the original key rate to the optimal key rate in the case of finite-size regime. It can be seen that with the decrease of the block length, both the secret key rate and the farthest transmission distance of the two models decrease. As shown in the Fig.~\ref{fig5}, the protocol can safely transmit more than 140 km, and when the block length is shortened to $N=10^9$, it can only safely transmit about 70 km. Moreover, the comparison of the secret key rate at $\eta_{BS}=0.5$ and key rate with the optimal $\eta_{BS}$ under different block lengths is also presented. For the optimal value of $\eta_{BS}$, there is a sightly increase in the secret key rate and even more obvious compared with the asymptotic-limit regime, and the problem caused by the deviation of heterodyne detection is worse. Therefor, the block length reduces the transmission distance of the secret key and has more effect on the optimization degree of the key rate.

The relationship between tolerable excess noise of the modified model and transmission distance with different block lengths is presented in Fig.~\ref{fig9}. Obviously, the noise tolerance of the model decreases compared with that of asymptotic-limit regime. With the increase of transmission distance and the decrease of the block length, the value of tolerable noise gradually decreases, which furthermore decreases to 0 at 90 km, 140 km and 190 km respectively, while the minimum tolerable excess noise is about 0.05 at the optimal modulation variance in the asymptotic-limit regime.

Figure~\ref{fig10} illustrates the impact of modulation variance on the secret key rate, presenting a typical scenario with a transmission distance of $L$ = 80 km. It can be observed that the key rate initially increases with the increase in modulation variance, reaching an optimal modulation variance before decreasing with further increases in modulation variance. From the simulation results, the available range of modulation variances that can guarantee a positive key rate is also influenced by the length of the data block. When the block length is shorter, the allowable range of modulation variances to achieve a positive key rate is narrower compared to cases when the block length is longer. The limited nature of the block also constrains the choice of modulation variance.

\section{\label{sec:level5}Conclusion}
In this paper, a modified biased No-Switching protocol is proposed, which focuses on closing the practical security loophole in the experiment and improves the secret key rate. 
The security loophole in a practical system mainly comes from the mismatch between the ideal protocol and the practical system caused by the asymmetry in heterodyne detection. Experimental results of the beam splitter and two homodyne detectors are carried out to verify the asymmetry. It can be found that the actual transmittance of beam splitter is not accurate $50:50$, and the output of the two homodyne detectors is different. The modified biased No-Switching protocol avoids this problem by calibrating the two homodyne detectors separately, which can be more compatible with the current practical system. Moreover, an optimization strategy is proposed to improve the secret key rate by adjusting the transmittance of the beam splitter. The security of the modified biased No-Switching protocol is analyzed in detail, and the simulation results of both the conventional detector model and the modified detector model are provided, with the comparison of secret key rate and performance between the two models.

According to the simulation results, the secret key rate of the practical biased No-Switching protocol is to some extent lower than that of the ideal No-Switching protocol, which indicates that neglecting the asymmetry in heterodyne detection will lead to an overestimation of the secret key rate and introduce a practical security loophole. For both detector models, the performances are approximately identical, with the key rate of the conventional model slightly higher than that of the modified model, suggesting that the modified detector model can simplify the practical detector modeling process according to the actual situation and hardly affect the performance of the protocol. By calibrating the transmittance $\eta_{BS}$ of the front beam splitter in front of the heterodyne detector, the key rate of protocol has been successfully optimized. From the comparison, the optimized key rates are improved at all distances, with a more pronounced improvement in the finite-size regime, which indicates that the protocol is more close to the practical application. The modified biased No-Switching protocol significantly improves the practical security and optimizes the secret key rate, which meets the requirements on security and high-performance of a practical system.

\section{\label{sec:level6}ACKNOWLEDGMENTS}
This research was supported by the National Natural Science Foundation of China (62001044), the Basic Research Program of China (JCKY2021210B059), the Equipment Advance Research Field Foundation (315067206), and the Fund of State Key Laboratory of Information Photonics and Optical Communications. 
\appendix
\nocite{*}


\bibliography{apssamp.bib}
\end{document}